\documentclass[preprint,amsmath,amssymb,superscriptaddress, floatfix,nofootinbib]{revtex4}
\usepackage{epsfig,subfigure}
\usepackage{epstopdf}
\usepackage[utf8]{inputenc}
\usepackage{graphicx}
\usepackage{amssymb}
\usepackage{amsxtra}
\usepackage{amsmath}
\usepackage{booktabs,multirow,tabularx}
\usepackage{slashed}
\usepackage{float}
\usepackage{placeins}
\usepackage{rotating}
\usepackage{lscape}
\usepackage{color}
\usepackage{hyperref}
\usepackage{soul}
\usepackage{cleveref}

\crefformat{equation}{Eq.~(#2#1#3)}

\newcommand{\bea}{\begin{eqnarray}}
\newcommand{\eea}{\end{eqnarray}}

\def\beq{\begin{equation}}
\def\eeq{\end{equation}}

\DeclareUnicodeCharacter{2212}{-}

\begin{document}
\title{Electroweak~Restoration~at~the~LHC~and~Beyond:~The~$Vh$~Channel\vspace{-0.1in}}

\author{Li Huang}
\email{huangli@ku.edu}
\affiliation{Department of Physics and Astronomy,~ University of Kansas,~ Lawrence,~ KS 66045~ USA}

\author{Samuel D. Lane}
\email{samuel.lane@ku.edu}
\affiliation{Department of Physics and Astronomy,~ University of Kansas,~ Lawrence,~ KS 66045~ USA}
\affiliation{Department of Physics, Brookhaven National Laboratory, Upton, NY 11973~ U.S.A.}

\author{Ian M. Lewis}
\email{ian.lewis@ku.edu}
\affiliation{Department of Physics and Astronomy,~ University of Kansas,~ Lawrence,~ KS 66045~ USA}

\author{Zhen Liu}
\email{zliuphys@umd.edu}
\thanks{\scriptsize \!\! \href{https://orcid.org/0000-0002-3143-1976}{0000-0002-3143-1976}}
\affiliation{Maryland Center for Fundamental Physics, Department of Physics, University of Maryland, College Park, MD, 20742 ~ USA}
\affiliation{School~of~Physics~and~Astronomy,~University~of~Minnesota,~Minneapolis,~MN,~55455~USA}


\begin{abstract}
The LHC is exploring electroweak (EW) physics at the scale EW symmetry is broken. As the LHC and new high energy colliders push our understanding of the Standard Model to ever-higher energies, it will be possible to probe not only the breaking of but also the restoration of EW symmetry. We propose to observe EW restoration in double EW boson production via the convergence of the Goldstone boson equivalence theorem. This convergence is most easily measured in the vector boson plus Higgs production, $Vh$, which is dominated by the longitudinal polarizations. We define EW restoration by carefully taking the limit of zero Higgs vacuum expectation value (vev). EW restoration is then measured through the ratio of the $p_T^h$ distributions between $Vh$ production in the Standard Model and Goldstone boson plus Higgs production in the zero vev theory, where $p_T^h$ is the Higgs transverse momentum. As EW symmetry is restored, this ratio converges to one at high energy. We present a method to extract this ratio from collider data. With a full signal and background analysis, we demonstrate that the 14 TeV HL-LHC can confirm that this ratio converges to one to 40\% precision while at the 27 TeV HE-LHC the precision will be 6\%. We also investigate statistical tests to quantify the convergence at high energies.  Our analysis provides a roadmap for how to stress test the Goldstone boson equivalence theorem and our understanding of spontaneously broken symmetries, in addition to confirming the restoration of EW symmetry.
\end{abstract}

\maketitle

\section{Introduction}
\label{sec:intro}
With the discovery of a Standard Model-like Higgs boson~\cite{Aad:2012tfa,Chatrchyan:2012ufa}, we entered a new era of probing the nature of electroweak (EW) symmetry breaking. Through the measurement of many EW and Higgs boson processes, the LHC is exploring the nature of a spontaneously broken symmetry at and above its breaking scale. As the LHC continues to gather data, it pushes these precision measurements and our understanding of EW symmetry breaking (EWSB) to ever-higher energies. These higher energies are very interesting for precision EW measurements~\cite{Mangano:2016jyj,Farina:2016rws,Englert:2016hvy,Green:2016trm,Butter:2016cvz,Zhang:2016zsp,Franceschini:2017xkh,Brivio:2017vri,Baglio:2017bfe,Liu:2018pkg,Baglio:2018bkm,Almeida:2018cld,Ellis:2018gqa,Biekotter:2018jzu,Alves:2018nof,DiLuzio:2018jwd,Banerjee:2018bio,Hays:2018zze,Biekotter:2018rhp,Grojean:2018dqj,Henning:2019vjr,deBlas:2019wgy,Banerjee:2019twi,Cuomo:2019siu,Chiu:2019ksm,Brehmer:2019gmn,Azzi:2019yne,Baglio:2019uty,Delgado:2019tbz,Freitas:2019hbk,Bishara:2020vix,Ricci:2020xre,Bishara:2020pfx,Araz:2020zyh,Baglio:2020oqu,Dawson:2020oco,Ellis:2020unq,Aoude:2020dwv}, in particular at future high energy colliders. As we get further above the EW scale, EW particles are essentially massless, and new interesting SM physics begins to appear. For example, the massive EW gauge bosons become partons and must be included in parton distribution functions as EW multiplets~\cite{Dawson:1984gx,Kane:1984bb,Kunszt:1987tk,Bauer:2017isx,Bauer:2018arx,Han:2020uid} and parton showers~\cite{Hook:2014rka,Chen:2016wkt,Bauer:2018xag,Cuomo:2019siu}. While these effects are intrinsically interesting and necessary to our understanding of the SM, they have also been shown to significantly impact searches for beyond the SM physics~\cite{Rinchiuso:2020skh}.

In this paper, we propose a new study to test one of the central behaviors of the SM at high energy: restoration of EW symmetry. We will propose a systematic analysis to observe this restoration at high energy colliders. Our study will open new analysis methods to stress test our understanding of the SM and the spontaneous breaking of EW theory. The main ingredient of our analysis is that as massive bosons become massless, their longitudinal modes can be replaced by the associated Goldstone bosons via the Goldstone boson equivalence theorem (GBET). Indeed, the GBET is a central ingredient to our understanding of the quantum field theory of spontaneously broken symmetries. Hence, our analysis provides a roadmap for how to empirically test the GBET, deepen our knowledge of spontaneously broken symmetries, and confirm the restoration of EW symmetry at high energies.

In the SM, restoration of EW symmetry is equivalent to taking the limit where the Higgs vacuum expection value (vev), $v$, goes to zero~\footnote{In a consistent manner.}.  In this limit the EW gauge bosons become massless.  That is, only the transverse polarizations persist and the longitudinal polarizations are replaced by their associated Goldstone bosons, i.e. the GBET mentioned above. There is a long history~\cite{LlewellynSmith:1973yud,Veltman:1976rt,Lee:1977eg,Chanowitz:1985hj,Dicus:1990fz,Barger:1990py,Bagger:1992vu,Dicus:1992vj,Bagger:1993zf,Bagger:1995mk,Chanowitz:1998wi,Butterworth:2002tt,Han:2009em,Brehmer:2014pka} of trying to observe the GBET via longitudinal vector boson scattering.  One of the interesting things about longitudinal vector boson scattering is that this process probes the quartic Goldstone boson coupling, which arises via the Higgs potential:
\begin{eqnarray}
V(H)=-\mu^2 H^\dagger H+\lambda\left(H^\dagger H\right)^2\quad{\rm where}\quad H=\begin{pmatrix} G^+\\ \frac{1}{\sqrt{2}}\left(v+h+i\,G^0\right)\end{pmatrix}\label{eq:Vpot}
\end{eqnarray}
is the Higgs doublet, $G^+,G^0$ are the Goldstone bosons, $h$ is the Higgs boson, and $v=246$~GeV is the Higgs vev.  Hence, longitudinal vector boson scattering probes the shape of the Higgs potential and the source of EWSB.  Additionally, this process violated perturbative unitarity without a Higgs boson~\cite{Veltman:1976rt,Lee:1977eg,Chanowitz:1985hj}.  However, with the observation of a light Higgs boson with SM-like couplings to EW gauge bosons~\cite{Khachatryan:2016vau,deFlorian:2016spz,Sirunyan:2018koj,Aad:2019mbh}, longitudinal vector boson scattering is effectively unitarized with the violation of perturbative unitarity pushed to multi-TeV energies~~\cite{Lee:1977eg,Belyaev:2012bm,Brehmer:2014pka,Corbett:2017qgl,Ballestrero:2017bxn,Falkowski:2019tft,Chang:2019vez}, making it difficult to observe.

As the above makes clear, the observation of EW symmetry restoration and the GBET is simplest in processes that are dominated by longitudinally polarized gauge bosons.  Such a process is Higgs production in association with an EW gauge boson: $q\bar{q}'\rightarrow Vh$ with $V=W^\pm,Z$ ($Vh$).  In the GBET, the $q\bar{q}'\rightarrow Vh$ production is equivalent to $q\bar{q}'\rightarrow G^{\pm,0}h$ production ($Gh$) which arises from the Higgs kinetic term:
\begin{eqnarray}
\mathcal{L}_{\rm kin}=|D_\mu H|^2.
\end{eqnarray}
The kinetic term contains the trilinear interactions (a) $Z-G^0-h$, $W^\pm-G^\mp-h$ and (b) $Z/\gamma-G^+-G^-$, $W^\pm -G^\mp-G^0$.  The interactions (a) contribute to the processes $q\bar{q}\rightarrow V_L h$, where the subscript $L$ indicates a longitudinally polarized vector boson.  The interactions (b) contribute to pair production of longitudinally polarized gauge bosons $q\bar{q}\rightarrow V_L V'_L$, where $V'=Z,W^\pm$.  However, the pair production of gauge bosons $q\bar{q}'\rightarrow VV'$ is dominated by transverse polarizations to high energy~\cite{Baglio:2017bfe,Baglio:2018rcu,Denner:2020bcz}. For the $Vh$ channel, the contribution from transversely polarized vector bosons is suppressed since a portion of the Higgs doublet already exists in the final state.

From this discussion, Higgs production in association with $W^\pm$ or $Z$ is a prime candidate to observe EW restoration.  In this paper we present an analysis strategy to do precisely this.  While this may seem straightforward, complications immediately arise when trying to observe EW restoration at hadron colliders.  Namely, the vector and Higgs bosons are intermediate states, and the collider observes their decay products.  These decays occur at the EW scale and the GBET is not valid.  This is clear by noting that while vector boson couplings to fermions are universal across generations, the Goldstone bosons couple like mass.  Hence, their branching ratios are vastly different and it is necessary to unfold to the underlying two-to-two process.  

As a proof of principle, we show that the convergence of the EW restoration can be observed in $q\bar{q}'\rightarrow Vh$ in the Higgs transverse momentum distribution at the 14 TeV high luminosity LHC (HL-LHC) and the proposed 27 TeV high energy LHC (HE-LHC).  We will define a signal strength that is a ratio of the vector boson $Vh$ and Goldstone boson $Gh$ processes.  In fact, numerically the signal strength is the same for both $W^\pm h$ and $Zh$ production.  At high Higgs transverse momentum, we show that it is possible to observe the convergence of this signal strength to one, indicating that EW symmetry is restored.  We will also explore various test statistics to determine how well the GBET converges.  In particular, we propose a modified the Kullback-Leibler divergence to quantify the convergence.

In Sec.~\ref{sec:Theory} we give the theoretical foundation for our work.  Helicity amplitudes of di-boson processes and polarized production rates are given in Sec.~\ref{sec:Theory}(a), and in Sec.~\ref{sec:Theory}(b) we define what we mean by the $v\rightarrow 0$ limit.  In Sec.~\ref{sec:likelihood} we define a likelihood to perform the unfolding, and define the relevant signal strength.  We present our collider analysis in Sec.~\ref{sec:collider}, which is based on a deep neural network (DNN).  In Sec.~\ref{sec:results}, we present our results showing the convergence of our signal strength as well as the modified Kullback-Leibler divergence.  Finally, in Sec.~\ref{sec:conc} we conclude.

\section{Theory}
\label{sec:Theory}
\subsection{Amplitudes}
To observe the convergence of the Goldstone boson equivalence theorem and restoration of EW symmetry, we need to look at EW gauge boson processes that are longitudinally dominated at high energy.  To determine the channels to study, we first calculate di-boson helicity amplitudes in the high energy limit~\cite{Gaemers:1978hg,Duncan:1985vj,Hagiwara:1986vm,Baglio:2018rcu}.  The fully longitudinal double EW gauge boson production modes are
\begin{eqnarray}
\mathcal{A}(q_+\bar{q}_-\rightarrow W^+_LW^-_L)&=&-i\,\frac{e^2\,Q_q}{2\, c_W^2}\sin\theta+\mathcal{O}(\hat{s}^{-1})\,,\nonumber\\
\mathcal{A}(q_-\bar{q}_+\rightarrow W^+_LW^-_L)&=&i\,\frac{e^2\,T_3^q}{6\,c_W^2\,s_W^2}\left(3\,c_W^2+2\,T_3^q\,s_W^2\right)\sin\theta+\mathcal{O}(\hat{s}^{-1})\,,\label{eq:VVlong}\\
\mathcal{A}(q_-\bar{q}'_+\rightarrow W^\pm_LZ_L)&=&-i\,\frac{e^2\,T_3^q}{\sqrt{2}s_W^2}\sin\theta+\mathcal{O}(\hat{s}^{-1})\,,\nonumber\\
\mathcal{A}(q_\pm \bar{q}'_\mp\rightarrow Z_L Z_L)&=&\mathcal{O}(\hat{s}^{-1})\,,\nonumber
\end{eqnarray}
where $\sqrt{\hat{s}}$ is the partonic center of mass energy, the subscript $L$ on EW gauge bosons indicates longitudinal polarization, and the subscripts on the quarks indicate quark helicity. For $W^+W^-$ production $\theta$ is the angle between $W^+$ and initial state quark, and for $WZ$ production $\theta$ is the angle between the $W$ and the initial state quark. 
$Q_q$ is the quark $q$'s charge, $T_3^q$ is the quark $q$'s isospin, and $c_W=\cos\theta_W,\,s_W=\sin\theta_W$ is the weak mixing angle. 

As expected, the fully longitudinal EW gauge boson pair production modes $W^+W^-$ and $WZ$ persist at high energy. However, so do transversely polarized gauge bosons with opposite helicities:
\begin{eqnarray}
\mathcal{A}(q_-\bar{q}_+\rightarrow W^+_\pm W^-_\mp)&=&\mp i \frac{e^2}{2\,s_W^2}\frac{1+2\,T_3^q\,\cos\theta}{1\pm\cos\theta}\sin\theta+\mathcal{O}(\hat{s}^{-1})\,,\label{eq:VVtrans}\\
\mathcal{A}(q_-\bar{q}'_+\rightarrow W^\pm_\pm Z_\mp)&=&\mp i\frac{e^2}{\sqrt{2}\,s_W^2\,c_W}\left(g_L^{q'Z}(1+\cos\theta)+g_L^{qZ}(1-\cos\theta)\right)\frac{\sin\theta}{1\pm \cos\theta}+\mathcal{O}(\hat{s}^{-1})\,,\nonumber\\
\mathcal{A}(q_-\bar{q}_+\rightarrow Z_+ Z_-)&=&2\,i\,\frac{e^2}{s_W^2c_W^2}{g_L^{qZ}}^2\sqrt{\frac{1-\cos\theta}{1+\cos\theta}}+\mathcal{O}(\hat{s}^{-1})\,,\nonumber\\
\mathcal{A}(q_+\bar{q}_-\rightarrow Z_+ Z_-)&=&-2\,i\,\frac{e^2}{s_W^2c_W^2}{g_R^{qZ}}^2\sqrt{\frac{1+\cos\theta}{1-\cos\theta}}+\mathcal{O}(\hat{s}^{-1})\,,\nonumber
\end{eqnarray}
where the subscript $\pm$ on EW gauge bosons indicate the transverse helicities, for $ZZ$ final state $\theta$ is the angle between the initial state quark and $Z_+$, and
\begin{eqnarray}
g_L^{qZ}=T_3^q-Q_q\,s_W^2,\quad{\rm and}\quad g_R^{qZ}=-Q_q\,s_W^2.
\end{eqnarray}
All other amplitudes are either zero or suppressed at high energies:
\begin{gather}
\mathcal{A}(q_\pm\bar{q}_\mp\rightarrow W^\pm_\pm W^\mp_L)\sim\mathcal{A}(q_-\bar{q}'_+\rightarrow W^\pm_\pm Z_L)\sim\mathcal{A}(q_-\bar{q}'_+\rightarrow Z_\pm W^\pm_L)\sim\mathcal{A}(q_\pm\bar{q}_\mp\rightarrow Z_\pm Z_L)\sim\mathcal{O}(\hat{s}^{-1/2})\,,\nonumber\\
\mathcal{A}(q_\pm\bar{q}_\mp\rightarrow W^+_\pm W^-_\pm)\sim\mathcal{A}(q_-\bar{q}'_+\rightarrow W^\pm_\pm Z_\pm)\sim\mathcal{A}(q_\pm\bar{q}_\mp\rightarrow Z_\pm Z_\pm)\sim\mathcal{O}(\hat{s}^{-1})\,,\nonumber\\
\mathcal{A}(q_+\bar{q}_-\rightarrow W^+_\pm W^-_\mp)=\mathcal{A}(q_+\bar{q}'_-\rightarrow W^\pm_\lambda Z_{\lambda'})=0.
\end{gather}

\begin{figure}[tb]
\begin{center}
\subfigure[]{\includegraphics[width=0.45\textwidth,clip]{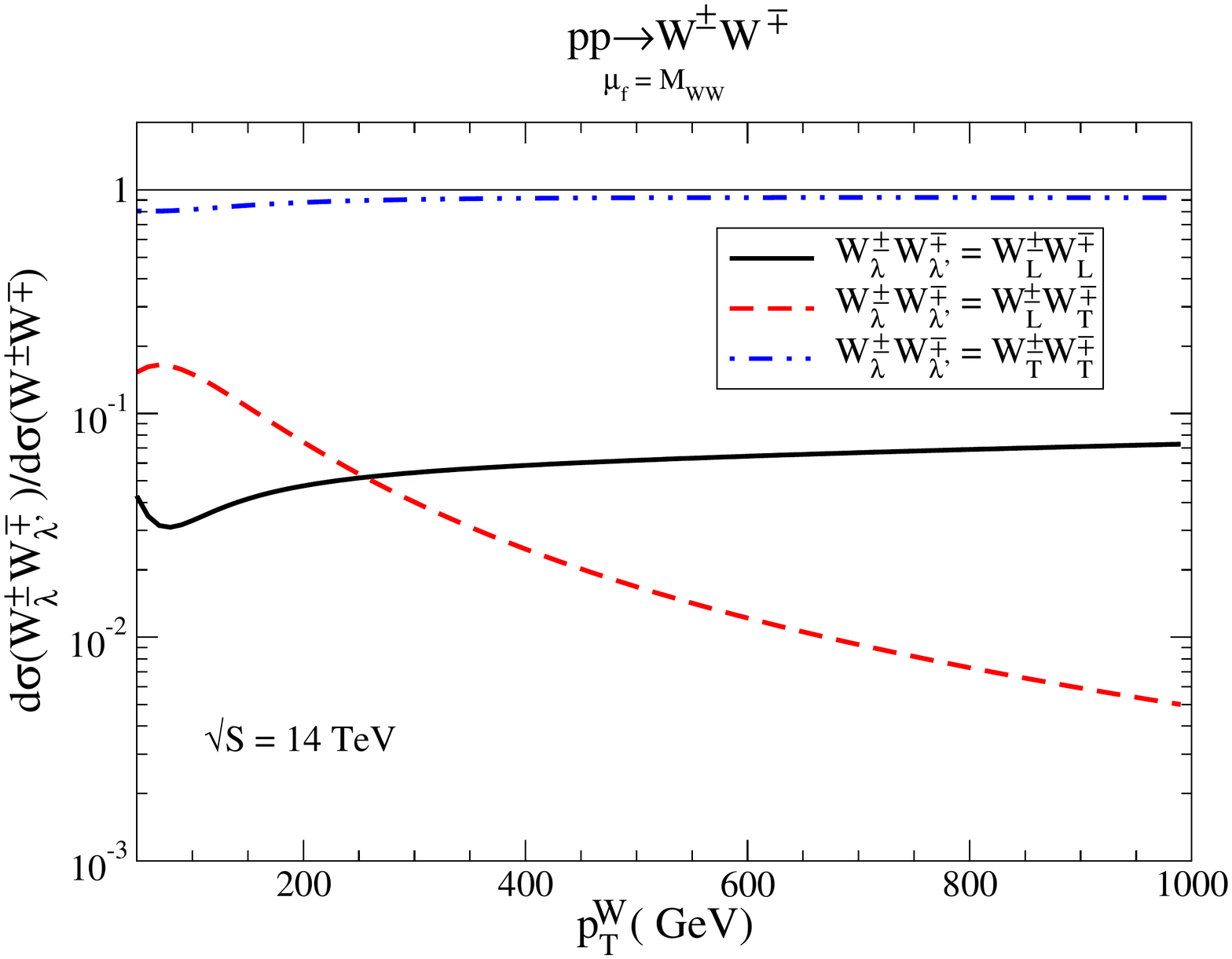}}
\subfigure[]{\includegraphics[width=0.45\textwidth,clip]{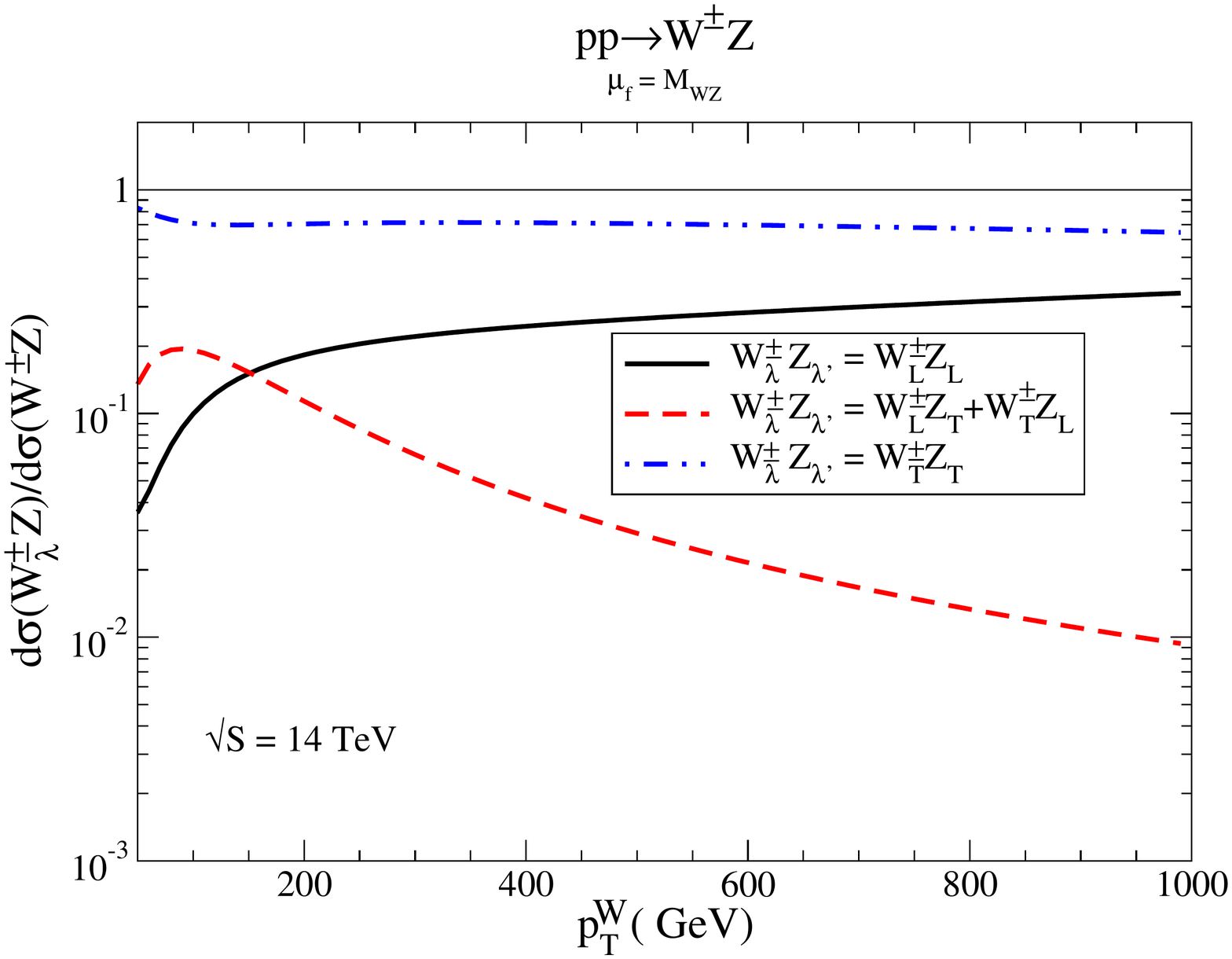}}
\subfigure[]{\includegraphics[width=0.45\textwidth,clip]{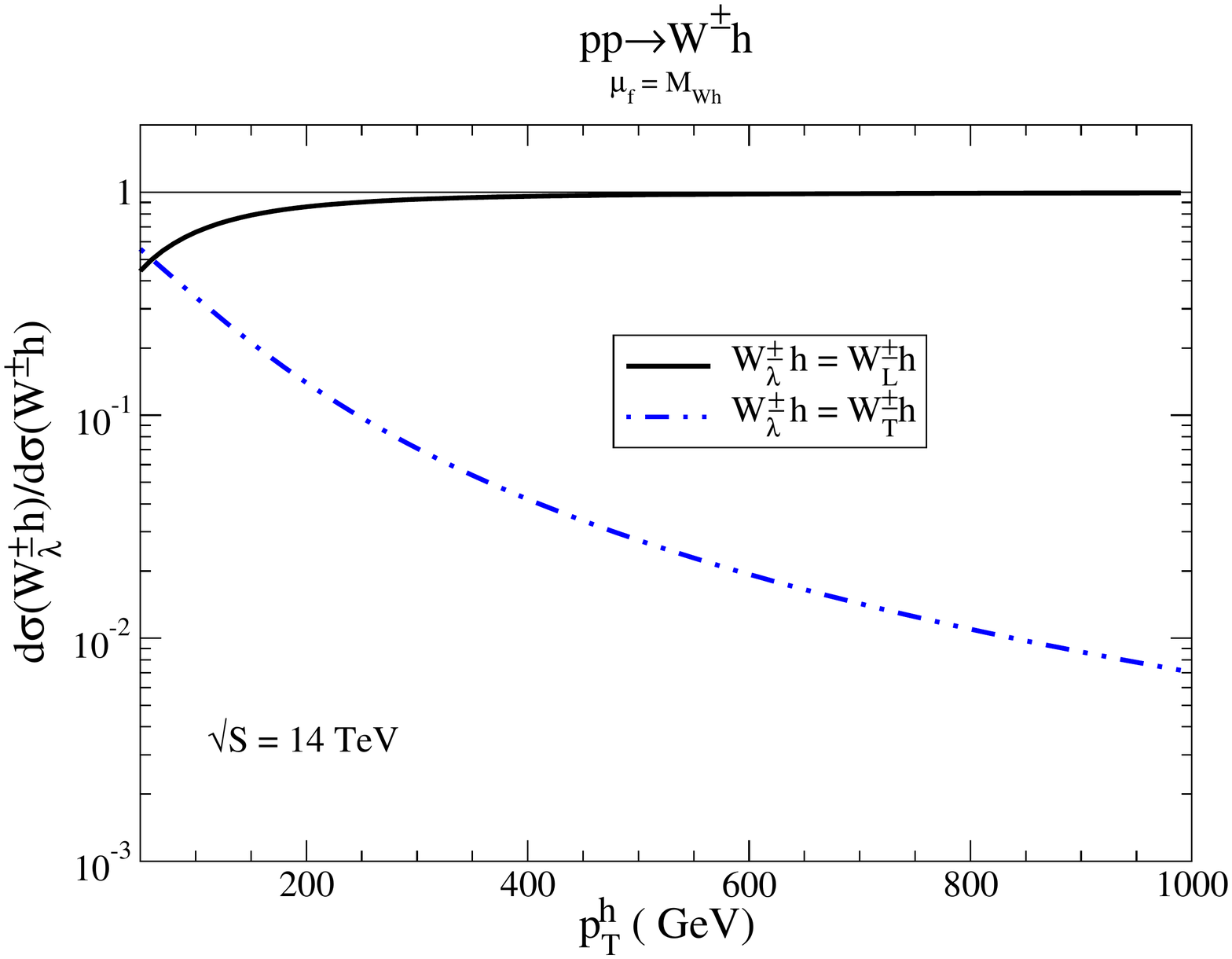}}
\subfigure[]{\includegraphics[width=0.45\textwidth,clip]{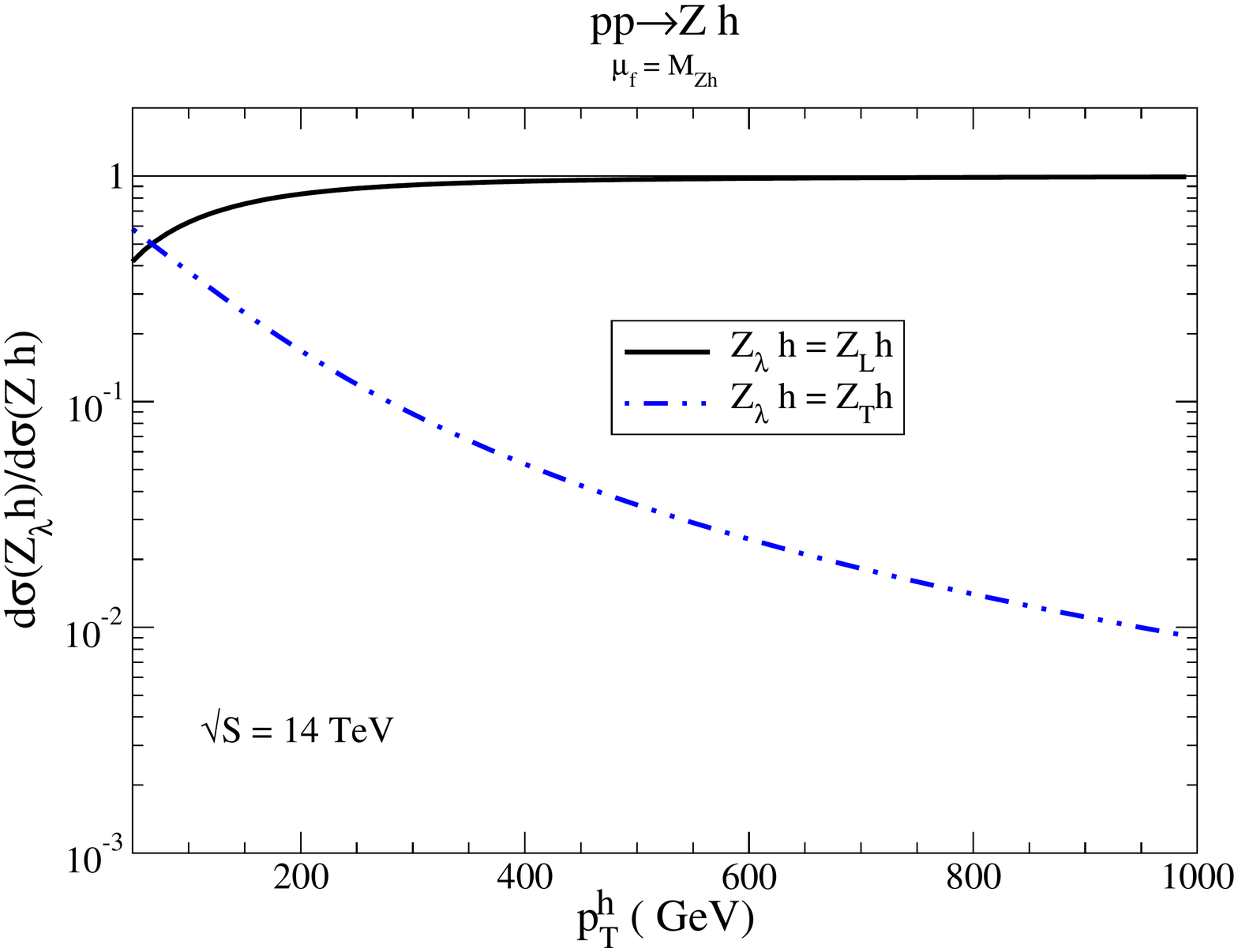}}

\end{center}
\caption{\label{fig:pT_Rat} Ratio of transverse momentum distributions of polarized gauge boson production to the total distribution summed over polarizations.  These are shown using $W$ transverse momentum, $p_T^W$, for (a) $W^\pm W^\mp$ and (b) $W^\pm Z$; and Higgs transverse momentum, $p_T^h$, for (c) $W^\pm h$ and (d) $Zh$.  The gauge boson polarizations are (blue dash-dot-dot) fully longitudinal, (black solid) fully transverse, and (red dashed) longitudinal+transverse.   The lab frame energy is the HL-LHC energy of $\sqrt{S}=14$~TeV.  The subscript $T$ on the gauge bosons indicate summed over transverse polarizations.}
\end{figure}

From Eqs.~(\ref{eq:VVlong}) and~(\ref{eq:VVtrans}) it is clear that double EW gauge boson production is not longitudinally dominated.  Indeed,  even though both fully longitudinal and transverse polarizations persist at high energy, as shown in Fig.~\ref{fig:pT_Rat}(a,b) $WW$ and $WZ$ production are strongly dominated by the transverse polarizations.  Here we use \texttt{CTEQ6L1} parton distribution functions (pdfs)~\cite{Pumplin:2002vw} implemented in \texttt{LHAPDF}~\cite{Buckley:2014ana} via \texttt{ManeParse}~\cite{Clark:2016jgm}.  This is particularly striking in $WW$ production where at high energies $90-95\%$ of the $W$s are transversely polarized, while $WZ$ production is $60-70\%$ transversely polarized.  Hence, to find the longitudinally polarized signal and observe EW restoration in $q\bar{q}'\rightarrow VV'$, either the differences in the angular distributions of the gauge bosons must be exploited or their polarizations must be tagged, which is very difficult~\cite{Duncan:1985vj,Azatov:2017kzw,Panico:2017frx,Liu:2018pkg,Aoude:2019cmc,De:2020iwq,Denner:2020bcz,Grossi:2020orx,Kim:2021gtv}.  There is also an additional complication that the gauge bosons are not final state particles and different gauge boson polarizations interfere with each other~\cite{Hagiwara:1986vm,Azatov:2017kzw,Panico:2017frx,Aoude:2019cmc,Denner:2020bcz}. 

These complications do not arise in EW gauge boson production in association with a Higgs~\cite{Banerjee:2018bio}:
\begin{eqnarray}
\mathcal{A}(q_+\bar{q}_{-}\rightarrow Z_Lh)&=&\pm\,i\,\frac{e^2\,g_{R}^{qZ}}{2\,c_W^2\,s_W^2}\sin\theta+\mathcal{O}(\hat{s}^{-1}),\label{eq:Vhlong}\\
\mathcal{A}(q_-\bar{q}_{+}\rightarrow Z_Lh)&=&\pm\,i\,\frac{e^2\,g_{L}^{qZ}}{2\,c_W^2\,s_W^2}\sin\theta+\mathcal{O}(\hat{s}^{-1}),\nonumber\\
\mathcal{A}(q_-\bar{q}'_+\rightarrow W_L^\pm h)&=&-i\,\frac{e^2}{2\,\sqrt{2}\,s_W^2}\sin\theta+\mathcal{O}(\hat{s}^{-1})\nonumber,\\
\mathcal{A}(q_\pm\bar{q}_{\mp}\rightarrow Z_\pm h)&\sim&\mathcal{A}(q_-\bar{q}'_+\rightarrow W_L^\pm h)\sim\mathcal{O}(\hat{s}^{-1/2}),\nonumber\\
\mathcal{A}(q_+\bar{q}'_-\rightarrow W^\pm_\pm h)&=&\mathcal{A}(q_+\bar{q}'_-\rightarrow W^\pm_\mp h)=0.\nonumber
\end{eqnarray}
The longitudinal polarizations persist at high energy while transverse polarizations decrease with energy.  This is even more clear in Figs.~\ref{fig:pT_Rat}(c,d), where the transverse polarizations make sub-percent level contributions to the total rate at high transverse momentum.  Since $q\bar{q}'\rightarrow Vh$ is quickly dominated by longitudinally polarized gauge bosons, there is no need to use polarization tagging to get a longitudinally enriched signal.  Hence, this channel is a prime candidate to observe EW restoration and the focus of our phenomenological analysis.

\subsection{EW Restoration}
To observe EW restoration, the $SU(2)\times U(1)$ symmetric phase with $v=0$ should be considered.  In this phase, the EW gauge bosons and SM fermions are massless.  To obtain $v=0$, the $\mu^2$ parameter in Eq.~(\ref{eq:Vpot}) must be zero or negative.  Hence, in principle the Higgs field could have a non-zero mass.  We will enforce the tree level relationships between the Higgs mass $m_h$, the $\mu^2$ parameter, and the vev then take the limit $v\rightarrow 0$:
\begin{eqnarray}
\mu^2&=&\lambda\,v^2\xrightarrow[v\rightarrow 0]{}0,\\
m_h^2&=&2\,\lambda\,v^2\xrightarrow[v\rightarrow 0]{}0.\nonumber
\end{eqnarray}
That is, we consider a massless Higgs doublet field consistent with the parameter relationships in the SM.

Once the $SU(2)\times U(1)$ symmetry is restored, calculations should be performed in the unbroken phase.  The relevant degrees of freedom are the $SU(2)$ gauge boson multiplet, the hypercharge gauge boson, the Higgs doublet, the left-handed fermion doublets, and the right-handed fermion singlets.  All fields are massless.  However, when considering collider phenomenology a couple complications arise.  First, vector bosons, Goldstones, and the Higgs boson are not final state particles.  Hence, their charges can be distinguished by the detector via their decay products.  This separates the components of the Higgs doublet and the $SU(2)\times U(1)$ gauge boson multiplets.  Second, each flavor of quark has a different pdf and the pdfs distinguish the components of the quark doublets.  Each of these effects break EW symmetry at the detector level.

With those considerations we compute Goldstone boson and Higgs production helicity amplitudes with initial and final states considered component-by-component.  For intermediate particles the massless gauge bosons of the unbroken $SU(2)\times U(1)$ are used.  The relevant helicity amplitudes for di-boson production are
\begin{eqnarray}
\mathcal{A}(q_+ \bar{q}_{-}\rightarrow G^0 h)&=&-\frac{e^2\,g_{R}^{qZ}}{2\,c_W^2\,s_W^2}\sin\theta,\label{eq:Gh}\\
\mathcal{A}(q_- \bar{q}_{+}\rightarrow G^0 h)&=&\frac{e^2\,g_{L}^{qZ}}{2\,c_W^2\,s_W^2}\sin\theta,\nonumber\\
\mathcal{A}(q_-\bar{q}_+\rightarrow G^\pm h)&=&\mp i \frac{e^2}{2\sqrt{2}s^2_W}\sin\theta,\nonumber\\
\mathcal{A}(q_-\bar{q}_+\rightarrow G^\pm G^0)&=&\frac{e^2}{2\,\sqrt{2}\,s_W^2}\sin\theta,\nonumber\\
\mathcal{A}(q_+\bar{q}_-\rightarrow G^+ G^-)&=&-i\,\frac{e^2Q_q}{2\,c_W^2}\sin\theta,\nonumber\\
\mathcal{A}(q_-\bar{q}_+\rightarrow G^+ G^-)&=&-i\,\frac{e^2\,T_3^q}{6\,c_W^2s_W^2}\left(3\,c_W^2+2\,T_3^qs_W^2\right)\sin\theta.\nonumber
\end{eqnarray}
As expected from the Goldstone boson equivalence theorem, the Goldstone boson production amplitudes agree with high energy longitudinal gauge boson amplitudes in Eqs.~(\ref{eq:VVlong}) and~(\ref{eq:Vhlong}).

\begin{figure}[tb]
\begin{center}
\includegraphics[width=0.45\textwidth,clip]{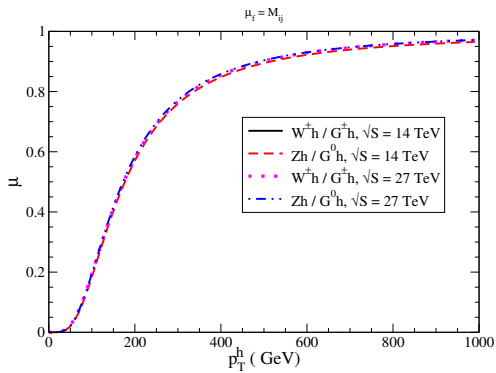}
\end{center}
\caption{\label{fig:sigstrength} Signal strengths (black solid) $\mu_{Wh}$, (red dashed) $\mu_{Zh}$ at $\sqrt{S}=14$~TeV and (magenta dotted) $\mu_{Wh}$, (blue dash-dot-dot) $\mu_{Zh}$ at $\sqrt{S}=27$~TeV.}
\end{figure}

To observe how quickly the Goldstone boson equivalence theorem converges in $Vh$ production, we define signal strengths as ratios of Higgs transverse momentum, $p_T^h$, distributions:
\begin{eqnarray}
\mu_{Wh}&=&\frac{d\sigma(pp\rightarrow W^\pm h)/dp_T^h}{d\sigma(pp\rightarrow G^\pm h)/dp_T^h},\nonumber\\
\mu_{Zh}&=&\frac{d\sigma(pp\rightarrow Zh)/dp_T^h}{d\sigma(pp\rightarrow G^0h)/dp_T^h}.\label{eq:sigstrength}
\end{eqnarray}
While $\sqrt{\hat{s}}$ is the relevant quantity for the convergence of the GBET, we use $p_T^h$ since it is more easily reconstructable when there is missing energy from gauge boson decays.  The signal strengths are shown in Fig.~\ref{fig:sigstrength} for both the HL-LHC with lab frame energy $\sqrt{S}=14$~TeV and the HE-LHC with $\sqrt{S}=27$~TeV.  While there is a very large difference between the $Vh$ and Goldstone boson plus Higgs distributions at low transverse momentum, they converge fairly quickly.  At transverse moment of $p_T^h\sim 400$~GeV, the $Vh$ and $Gh$ distributions agree at the $\sim 80\%$ level.  

Both $\mu_{Wh}$ and $\mu_{Zh}$ are in good agreement for the entire $p_T^h$ range at the HL- and HE-LHC.  Hence, a uniform signal strength can be defined for both $W^\pm h$ and $Zh$:
\begin{eqnarray}
\mu_{Vh}=\mu_{Wh}=\mu_{Zh}.~\label{eq:globalsigstrength}
\end{eqnarray}
Then both $W^\pm h$ and $Zh$ distributions can be fit to the same parameter, making the combination of these measurements straightforward.

\section{Signal Strength and Likelihood}
\label{sec:likelihood}
We now turn to how to observe EW restoration in the EW gauge boson plus Higgs boson production via the signal strength in Eq.~(\ref{eq:globalsigstrength}).   As discussed before, one immediate issue is that the vector and scalar bosons are not final state particles, and their decay products are detected.  The complication is that although the production of $Vh$ occurs at high energies, the decays of the vector boson and Higgs occur at the EW scale $\sim 100$~GeV where the Goldstone boson equivalence theorem is not a good approximation.

To extract the signal strength [Eq.~(\ref{eq:globalsigstrength})], the detector level events need to be unfolded~\cite{Cowan:2002in,Prosper:2011zz,Blobel:jla} to the partonic $q\bar{q}\rightarrow Vh$ level. There are many modern machine learning~\cite{Gagunashvili:2010zw,Glazov:2017vni,Datta:2018mwd,Andreassen:2019cjw} methods to unfold events.  However, we are primarily interested in only the Higgs transverse momentum distribution.  Hence, we adapt the unfolding method in Ref.~\cite{Sirunyan:2018kta} and use a simple one-dimensional likelihood function method.  For each bin of $p_T^h$ we define a likelihood function:
\begin{eqnarray}
\mathcal{L}_i(\Delta\sigma_1^{Vh},\Delta\sigma_2^{Vh},\ldots)=\frac{\left(\sum_j \Delta\sigma_j^{Vh}\epsilon_{ij}L+B_i\right)^{n_{obs,i}}}{n_{obs,i}!}e^{-\sum_j \Delta\sigma_j^{Vh}\epsilon_{ij}L-B_i},\label{eq:Li1}
\end{eqnarray}
where $i$ labels each $p_T^h$ bin; $B_i$ is the expected number of background events and $n_{obs,i}$ the total number of observed events in the $i$th bin; $L$ is the integrated luminosity; $\Delta\sigma_i^{Vh}$ is the partonic cross section in each bin; and $\epsilon_{ij}$ is an efficiency matrix.  The efficiency matrix takes into account detector effects, branching ratios, parton showering, and hadronization.  Using the uniform signal strength in Eq.~(\ref{eq:globalsigstrength}), the binned likelihood function is
\begin{align}
\mathcal{L}_i(\mu^1_{Vh},\mu^2_{Vh},\ldots)=\frac{\left(\sum_j \mu^j_{Vh}\Delta\sigma_j^{Gh}\epsilon_{ij}L+B_i\right)^{n_{obs,i}}}{n_{obs,i}!}e^{-\sum_j \mu^j_{Vh}\Delta\sigma_j^{Gh}\epsilon_{ij}L-B_i},\label{eq:Li}
\end{align}
where $\mu^j_{Vh}$ is the signal strength and $\Delta\sigma_j^{Gh}$ the Goldstone boson plus Higgs production rate in the $j$th Higgs transverse momentum bin.  For $Zh$ production the relevant Goldstone process is $G^0h$ and for $W^\pm h$ it is $G^\pm h$.  

The efficiency matrix takes care of the probability that a parton level event in the $i$th bin is in the $j$th bin at the detector level.  To calculate $\epsilon_{ij}$ we generate $Vh$ events in \texttt{MadGraph5\_aMC@NLO}~\cite{Alwall:2014hca} with parton showers and hadronization via \texttt{PYTHIA8}~\cite{Sjostrand:2014zea} and detector effects via \texttt{DELPHES3}~\cite{deFavereau:2013fsa}.  By comparing detector level reconstructed $p_T^h$ to the parton level information from~\texttt{MadGraph5\_aMC@NLO}, the efficiency matrix $\epsilon_{ij}$  can be determined.

Finally, we use the global likelihood across all bins
\begin{eqnarray}
\mathcal{L}(\mu^1_{Vh},\mu^2_{Vh},\ldots)=\prod_i \mathcal{L}_i(\mu^1_{Vh},\mu^2_{Vh},\ldots){\rm Pois}(n_{obs,i}|S_i+B_i),\label{eq:likelihood}
\end{eqnarray}
where ${\rm Pois}(x|y)$ is a conditional Poisson distribution and $S_i$ is the expected number of signal events in the $i$th bin.  Now, given a number of observed events $n_{obs,i}$, Eq.~(\ref{eq:likelihood}) is maximized to determine the binned signal strengths $\mu^i_{Vh}$.

\section{Collider Analysis}
\label{sec:collider}
We now turn to extracting our signal from background.  To get larger rates and clean signals, we consider $h\rightarrow bb$ and leptonic decays of the EW gauge boson.  Signal events are decomposed into six categories:
\begin{enumerate}
\item Two lepton final states, $Zh\rightarrow \ell^+\ell^- b\bar{b}$ , with either 
\begin{enumerate}
\item exactly two jets from the Higgs or
\item  three or more jets.
\end{enumerate}
\item One lepton final states, $Wh\rightarrow\ell\nu b\bar{b}$, with either 
\begin{enumerate}
\item exactly two jets from the Higgs or 
\item exactly three jets.
\end{enumerate}
\item Zero lepton final states,  $Zh\rightarrow\nu\nu b\bar{b}$, with either 
\begin{enumerate}
\item exactly two jets from $h\rightarrow bb$ or 
\item exactly three jets.
\end{enumerate}
\end{enumerate}
Note, for each signal with different multiplicities of jets, the efficiency matrix $\epsilon_{ij}$ in Eq.~(\ref{eq:Li}) must be recalculated to map onto the partonic $q\bar{q}\rightarrow Vh$ event.

The major backgrounds are: QCD production of $V+ll$, $V+$HF, $V+cl$ as well as top pair, single top and vector boson pair.  Here $l=u,\,d,\,s,\,g$, and HF indicates ``heavy flavor'': $bb,\,bc,\,cc,\,bl$.  For the zero and one-lepton signals, we include backgrounds from missing leptons.  The missing lepton rate is estimated by using the default setting of \texttt{DELPHES3}.

\subsection{Simulation}

We use \texttt{MadGraph5\_aMC@NLO}~\cite{Alwall:2014hca} for parton level generation and heavy particle decay, \texttt{PYTHIA8}~\cite{Sjostrand:2014zea} for parton showers and hadronization, \texttt{DELPHES3}~\cite{deFavereau:2013fsa} for fast detector simulation.   Finally, MLM jet matching~\cite{Mangano:2002ea} up to one additional jet is used for background simulation.

In the detector simulation, we use the default CMS card with some modifications.  While there is a \texttt{DELPHES3} card for Future Circular Colliders (FCC) that can be used for 27 TeV, for simplicity and direct comparison with 14 TeV results we use the CMS card for both the HL and HE-LHC. The FCC~\texttt{DELPHES3} card has electron tracking efficiency of 99\% for transverse momentum greater than 1 GeV.  Hence, we use the same basic acceptance cuts for leptons at both the 14 TeV HL-LHC and 27 TeV HE-LHC~\cite{Aaboud:2017xsd}:
\begin{itemize}
\item Lepton transverse momentum, $p_T^\ell$:\\
\begin{eqnarray}
p_T^\ell \geq 27~{\rm GeV}\,.
\end{eqnarray}
\item Lepton, $\eta_\ell$, and jet, $\eta_j$, rapidity:
\begin{eqnarray}
|\eta_\ell| \leq 2.5 ,\quad |\eta_j| \leq 5.0\,\label{eq:rap}
\end{eqnarray}
\item Minimum separation between jets, $j$, and leptons, $\ell$:
\begin{eqnarray}
\Delta R_{jj}\geq 0.4,\quad \Delta R_{j\ell}\geq 0.4,\quad \Delta R_{\ell\ell}\geq 0.4\,.
\end{eqnarray}
\item Electron isolation: \texttt{PTRatioMax}$=0.43$~\footnote{See Eq.~(3.1) of Ref.~\cite{deFavereau:2013fsa} for definition of \texttt{PTRatioMax}.}  considering particles with $p_T>0.5$~GeV and within a cone of radius $\Delta R<0.3$.
\item Muon isolation: \texttt{PTRatioMax}$=0.25$ considering particles with $p_T>0.5$~GeV and within a cone of radius $\Delta R<0.4$.
\end{itemize}
The minimum jet transverse momentum, $p_T^j$, requirement is different between 14 and {27 TeV}:
\begin{itemize}
\item At 14 TeV:
\begin{eqnarray}
p_T^j \geq 20~{\rm GeV}\,.\label{eq:pTj14}
\end{eqnarray}
\item At 27 TeV~\cite{Cepeda:2019klc}: 
\begin{eqnarray}
p_T^j \geq 30~{\rm GeV}\,.\label{eq:pTj27}
\end{eqnarray}
\end{itemize}
Finally, since our signal is rich in $b$-quarks, we also use a $b$ tagging rate of $0.70$ with mis-tag rates of $0.125$ for charm jets and $0.003$ for light jets~\cite{Aaboud:2018zhk}.

\subsection{Classification}

To classify signal from background, we use ``pre-cuts'' followed by a DNN.  The pre-cuts are basic multiplicity and invariant mass cuts to help separate signal and background:
\begin{itemize}
\item For the two lepton signals ($n_{\ell}=2$) we require exactly two same flavor, opposite sign leptons that reconstruct the $Z$ mass $|m_{\ell\ell}-m_Z|\leq 10$~GeV, where $m_{\ell\ell}$ is the di-lepton invariant mass.  In addition, we require at least two jets ($n_j\geq 2$) passing the cuts in Eqs.~(\ref{eq:rap},\ref{eq:pTj14},\ref{eq:pTj27})
\item For both the zero ($n_\ell=0$) and one lepton ($n_\ell=1$) signal we require either two or three jets ($n_j=2,3$) to pass the cuts in Eqs.~(\ref{eq:rap},\ref{eq:pTj14},\ref{eq:pTj27}).
\end{itemize}
For all signals we require exactly two $b$-tagged jets ($n_b=2$).

After events pass the pre-cuts, a DNN is used to further classify signal and background.  The inputs of the DNN are high-level reconstructed variables and are detailed in Appendix~\ref{app:DNN}.  The DNN is a binary classifier consisting of three hidden layers with $2^{10}$, $2^{12}$, and $2^{10}$ nodes.    We adopt \texttt{LeakyReLU}~\cite{Maas13rectifiernonlinearities} for non-linearity, use batch normalization between layers, and the output layer uses softmax to create a probability.  We use cross entropy as the loss function with an $L 2$ penalty:
\begin{eqnarray}
L=-y_s\log\,p-(1-y_s)\log(1-p)+\lambda \parallel W\parallel^2,
\end{eqnarray}
where $y_s$ is the signal indicator with $y_s=1$ for signal and $y_s=0$ for background, $p$ is the predicted signal probability, and $\parallel W\parallel^2$ is the matrix norm of the weight matrices.  While the same DNN structure is used for all six categories, the $L 2$ penalty value $\lambda$ changes.

\begin{table}[tb]
    \centering
\makebox[\textwidth][c]{
    \begin{tabular}{|l||c|c||c|c||c|c||c|c|}
    \hline
    \hline& \multicolumn{4}{c||}{14 TeV} & \multicolumn{4}{c|}{27 TeV}\\\cline{2-9}
 & \multicolumn{2}{c||}{$n_j=2$} & \multicolumn{2}{c||}{$n_j=3$}& \multicolumn{2}{c||}{$n_j=2$} & \multicolumn{2}{c|}{$n_j=3$}\\\cline{2-9}
                                       &Pre-Cut & DNN  &Pre-Cut & DNN&Pre-Cut & DNN  &Pre-Cut & DNN \\ \hline\hline
    $h_{bb}Z_{\ell\ell}$ &1.1 fb & 0.22 fb &1.1 fb &0.23 fb    &2.0 fb & 0.87 fb & 1.6 fb &1.2 fb \\ \hline\hline
    $Z$+HF &300 fb &1.4 fb & 530 fb &3.3 fb                    &580 fb &16 fb& 780 fb &120 fb \\ \hline 
    $tt$ &27 fb &0.14 fb & 69 fb &0.095 fb                          &92 fb &1.6 fb & 180 fb &19 fb \\ \hline
    single top &0.85 fb &0.0036 fb & 3.5 fb &0.0041 fb            &2.9 fb &0.047 fb & 11 fb &1.0 fb \\ \hline
    $Zcl$ &0.18 fb &0.0036 fb & 2.1 fb  &0.025 fb                   &0.75 fb&0.034 fb & 6.4 fb &0.94 fb\\ \hline
    $Zll$ &0.68 fb &0.019 fb & 13 fb  &0.20 fb                      &2.0 fb &0.096 fb & 27 fb &4.1 fb\\ \hline
    $VV'$ &4.8 fb  &0.026 fb &5.4 fb  &0.051 fb                    &6.5 fb  &0.22 fb &7.8 fb  &1.5 fb \\ \hline\hline
    Signal Significance & &9.4 & &6.5                          & &25 & &13 \\ \hline
    \hline
    \end{tabular}}
    \caption{Cut flow table and signal significance after the DNN for the two lepton categories.  The significances correspond to 3 ab$^{-1}$ at $14$~TeV and 15 ab$^{-1}$ at $27$ TeV. All backgrounds include possible decays leading to events with and without missing leptons.
    }
    \label{tab:2lep}
\end{table}

\begin{table}[tb]
    \centering
\makebox[\textwidth][c]{
    \begin{tabular}{|l||c|c||c|c||c|c||c|c|}
    \hline
    \hline& \multicolumn{4}{c||}{14 TeV} & \multicolumn{4}{c|}{27 TeV}\\\cline{2-9}
 & \multicolumn{2}{c||}{$n_j=2$} & \multicolumn{2}{c||}{$n_j=3$}& \multicolumn{2}{c||}{$n_j=2$} & \multicolumn{2}{c|}{$n_j=3$}\\\cline{2-9}
                                       &Pre-Cut & DNN  &Pre-Cut & DNN&Pre-Cut & DNN  &Pre-Cut & DNN \\ \hline\hline
    $h_{bb}W_{\ell\nu}$ &12 fb & 6.1 fb &7.3 fb & 0.38 fb      & 19 fb & 9.6 fb &9.8 fb & 1.2 fb \\ \hline\hline
    $W$+HF &580 fb &38 fb & 640 fb &0.035 fb                      &790 fb &43 fb & 940 fb &0.33 fb\\ \hline
    $Z$+HF &310 fb &8.5 fb &380 fb &$9.7\times10^{-5}$ fb                    &640 fb &21 fb & 670 fb &0.048 fb \\ \hline
    $tt$ &150 fb &15 fb & 560 fb &0.30 fb                        &580 fb &28 fb & 1500 fb &0.93 fb \\ \hline
    single top &11 fb &1.1 fb &68 fb &0.053 fb                   &36 fb &1.7 fb & 100 fb &0.12 fb \\ \hline
    $Wcl$ &4.9 fb &0.46 fb & 12 fb &$2.5\times10^{-3}$ fb                    &8.0 fb &0.56 fb & 19 fb &0.027 fb\\ \hline
    $Wll$ &10 fb &1.2 fb &36 fb &0.021 fb                        &28 fb &2.7 fb & 92 fb & 0.34 fb \\ \hline
    $Zcl$ &0.15 fb &$4.2\times10^{-3}$ fb  &0.51 fb & 0 fb                             &0.62 fb & 0.012 fb  &1.8 fb &$7.2\times10^{-5}$ fb \\ \hline
    $Zll$ &0.49 fb &0.014 fb &2.0 fb &$4.7\times10^{-5}$ fb                 &1.5 fb & 0.032 fb & 5.2 fb &$6.0\times10^{-4}$ fb \\ \hline
    $VV'$ &34 fb &2.0 fb &28 fb &0.015 fb                         &41 fb & 1.9 fb & 33 fb & 0.11 fb \\ \hline\hline
    Signal Significance & & 40 & &28                           & &120 & &98 \\ \hline
    \hline
    \end{tabular}}
    \caption{Cut flow table and signal significance after the DNN for the one lepton categories.  The significances correspond to 3 ab$^{-1}$ at $14$~TeV and 15 ab$^{-1}$ at $27$ TeV.  All backgrounds include possible decays leading to events with and without missing leptons.
    }
    \label{tab:1lep}
\end{table}

\begin{table}[tb]
    \centering
\makebox[\textwidth][c]{
    \begin{tabular}{|l||c|c||c|c||c|c||c|c|}
    \hline
    \hline& \multicolumn{4}{c||}{14 TeV} & \multicolumn{4}{c|}{27 TeV}\\\cline{2-9}
 & \multicolumn{2}{c||}{$n_j=2$} & \multicolumn{2}{c||}{$n_j=3$}& \multicolumn{2}{c||}{$n_j=2$} & \multicolumn{2}{c|}{$n_j=3$}\\\cline{2-9}
                                       &Pre-Cut & DNN  &Pre-Cut & DNN&Pre-Cut & DNN  &Pre-Cut & DNN \\ \hline\hline
    $h_{bb}Z_{\nu\nu}$ &9.8 fb
     & 4.7 fb &6.3 fb & 1.6 fb                                     &18 fb & 7.9 fb &9.6 fb & 1.4 fb \\ \hline\hline
    $W$+HF &310 fb &7.6 fb &440 fb &0.020 fb                        &420 fb & 14 fb &680 fb &0.028 fb \\ \hline
    $Z$+HF &2900 fb &110 fb &2900 fb &0.35 fb                       &5700 fb & 260 fb &5000 fb &0.72 fb\\ \hline
    $tt$ &7.6 fb &0.16 fb & 170 fb &0.041 fb                        &42 fb & 0.22 fb & 460 fb &0.020 fb\\ \hline
    single top &1.3 fb &0.035 fb & 22 fb &0.0091 fb                  &1.5 fb & 0.0057 fb & 19 fb &0.0019 fb\\ \hline
    $Wcl$ &1.1 fb &0.026 fb &4.2 fb &$5.3\times10^{-4}$ fb         &2.4 fb & 0.059 fb & 7.4 fb &0.0010 fb\\ \hline
    $Wll$ &3.7 fb &0.087 fb & 19 fb &0.014 fb                        &13 fb & 0.38 fb & 49 fb &0.028 fb\\ \hline
    $Zcl$ &1.4 fb &0.15 fb  &4.7 fb &0.0065 fb                      &3.3 fb & 0.23 fb  &9.0 fb &0.013 fb \\ \hline
    $Zll$ &6.8 fb &0.78 fb & 26 fb &0.12 fb                         &22 fb & 1.6 fb & 80 fb & 0.20 fb\\ \hline
    $VV'$ &68 fb &3.9 fb & 51 fb &0.084 fb                           &89 fb &4.7 fb & 65 fb &0.15 fb\\ \hline\hline
    Signal Significance & &23 & &84                                   & &58 & &140 \\ \hline
    \hline
    \end{tabular}}
    \caption{Cut flow table and signal significance after the DNN for the zero lepton categories.  The significances correspond to 3 ab$^{-1}$ at $14$~TeV and 15 ab$^{-1}$ at $27$ TeV.  All backgrounds include possible decays leading to events with and without missing leptons.
    }
    \label{tab:0lep}
\end{table}

Cut flow tables and signal significances are given in Tab.~\ref{tab:2lep} for  two lepton categories, Tab.~\ref{tab:1lep} for the one lepton categories, and Tab.~\ref{tab:0lep} for the zero lepton category at both 14 and 27 TeV.  The significances are calculated for the benchmark luminosities of 3 ab$^{-1}$ for the HL-LHC and 15 ab$^{-1}$ for the HE-LHC.  We use the asymptotic formula for a discovery significance with Poisson statistics
\begin{eqnarray}
\sigma=\sqrt{2 \left(\left(N_s+N_b\right)\times\log\left(1+\frac{N_s}{N_b}\right)-N_s\right)},
\end{eqnarray}
where $N_s,N_b$ are the number of signal and background events, respectively.  It is clear that background and signal are well separated.

\begin{figure}
\subfigure[]{\includegraphics[scale=0.45]{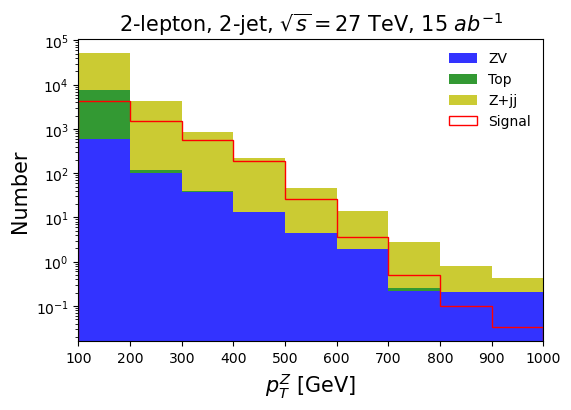}}
\subfigure[]{\includegraphics[scale=0.45]{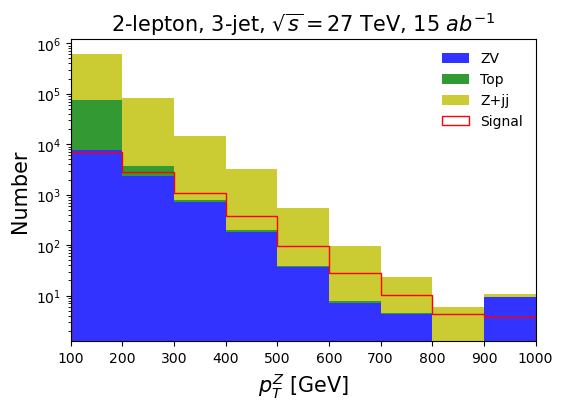}}
\subfigure[]{\includegraphics[scale=0.45]{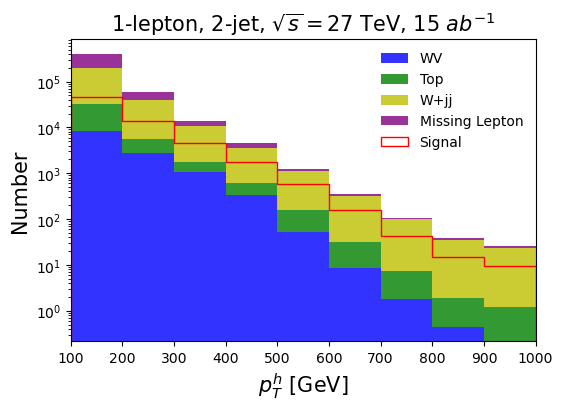}}
\subfigure[]{\includegraphics[scale=0.45]{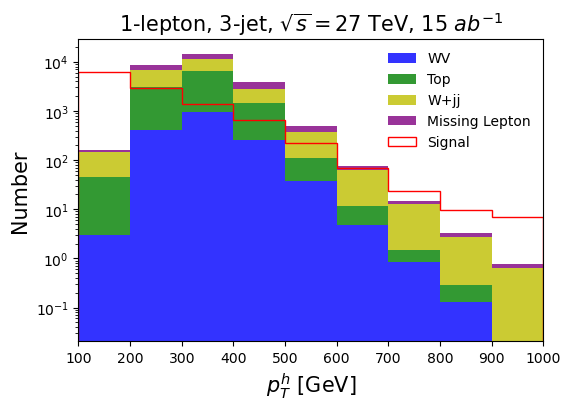}}
\subfigure[]{\includegraphics[scale=0.45]{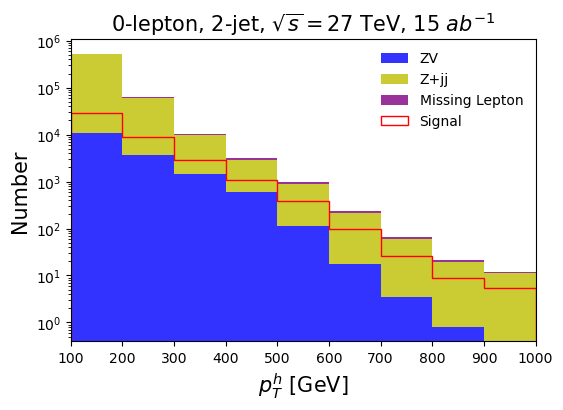}}
\subfigure[]{\includegraphics[scale=0.45]{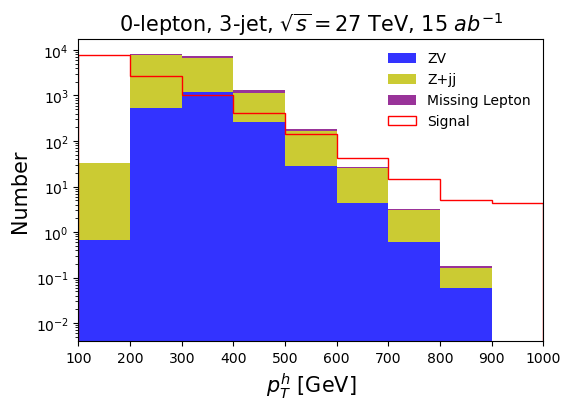}}
\caption{The number of events for (red line) signal and (colored bars) background at 27 TeV with 15 ab$^{-1}$ of data.  The background is cumulative and the signal overlaid.  We show (a,b) two lepton, (c,d) one lepton, and (e,f) zero lepton signal categories for (left hand side) two jet and (right hand side) three jet channels. The last bin is an overflow bin. $V+jj$ backgrounds include $V+HF$, $V+cl$, and $V+ll$.  Here, $ZV$ and $WV$ (Top) indicate $VV'$ ($tt$ and single top) backgrounds with no missing leptons.  ``Missing lepton'' indicates backgrounds where a lepton is missed, which is all other backgrounds except those explicitly listed.}
\label{resultAfterDNN}
\end{figure}

In Fig.~\ref{resultAfterDNN} we show the reconstructed vector boson $p_T$ distributions after the DNN selection for all six categories.  The background is cumulative, and the signal is overlaid.  At high energies the signal and background separation is better.  This is precisely where we expect to see EW restoration.

\section{Results}
\label{sec:results}
To fit the signal strengths in Eq.~(\ref{eq:globalsigstrength}) we perform pseudoexperiments to sample the binned $p_T^h$ distribution.  After the collider analysis of the previous section, we have a sample of signal and background events.  That sample is used to create a probability density function (PDF) for the signal and background $p_T^h$ distribution.  The total number of events is sampled according to a Gaussian distribution with the mean $\nu = S_{tot}+B_{tot}$ and standard deviation $\sigma=\sqrt{\nu}$, where the total number of expected signal and background events are
\begin{eqnarray}
S_{tot}=\sum_i S_i,\quad B_{tot}=\sum_i B_i,
\end{eqnarray} 
respectively, and $S_i, B_i$ are the expected number of signal and background events in the $i$th bin after the DNN, respectively.  The total number of events is then distributed according to the $p_T^h$ PDF.  In practice, instead of the Higgs transverse momentum, we use the di-lepton $p_T$ for two lepton categories.  At tree level, this is equivalent to $p_T^h$ for the $Vh$ signal.  For the zero and one lepton categories, we do use the reconstructed Higgs $p_T$.

\begin{figure}[htb]
\subfigure[]{\includegraphics[width=0.45\textwidth,clip]{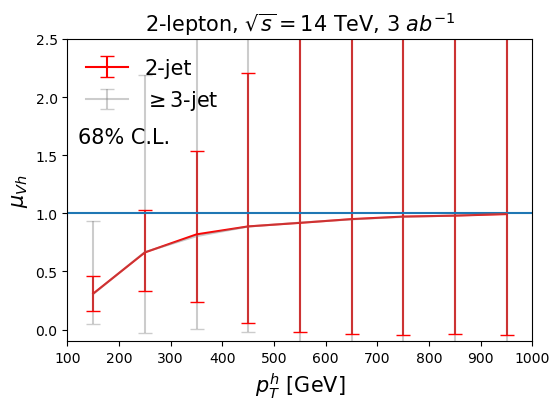}}
\subfigure[]{\includegraphics[width=0.45\textwidth,clip]{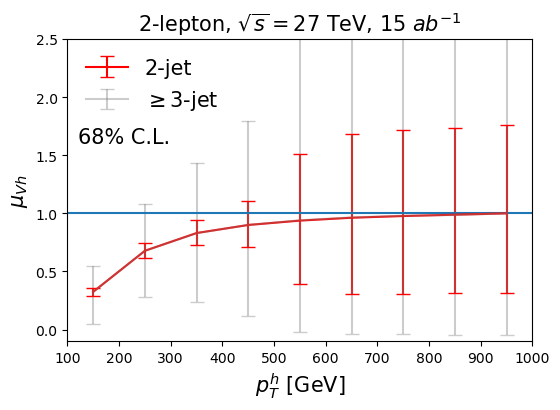}}
\subfigure[]{\includegraphics[width=0.45\textwidth,clip]{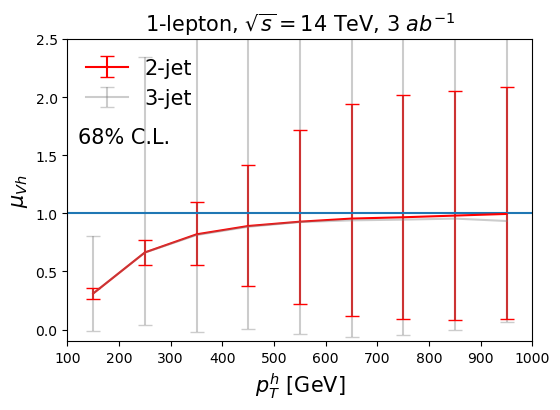}}
\subfigure[]{\includegraphics[width=0.45\textwidth,clip]{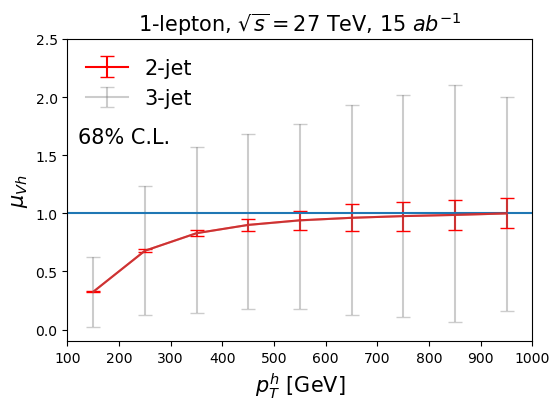}}
\subfigure[]{\includegraphics[width=0.45\textwidth,clip]{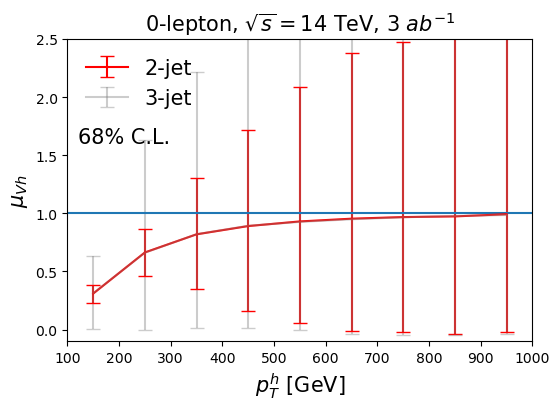}}
\subfigure[]{\includegraphics[width=0.45\textwidth,clip]{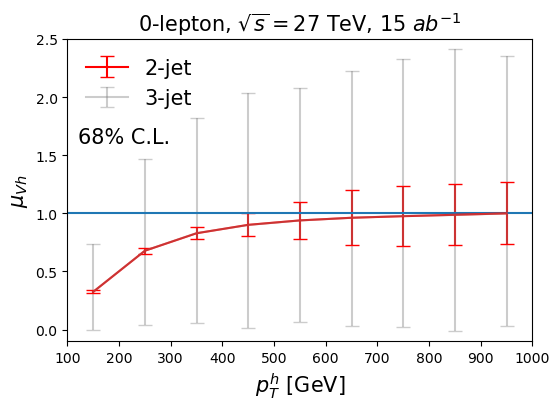}}
\caption{\label{fig:mu_cat} Signal strengths $\mu_{Vh}$ central values and 68\% CL for (a,b) two lepton categories, (c,d) one lepton categories, and (e,f) zero lepton categories for (a,c,e) 14 TeV with 3 ab$^{-1}$ and (b,d,f) 27 TeV with 15 ab$^{-1}$.  We show the (red) two jet and (gray) three jet categories separately.}
\end{figure}

For each of the six categories, we perform these pseudoexperiments at 14 TeV and 27 TeV.  Then the $p_T^h$ distribution is repeatedly sampled for each pseudoexperiment.  These samples determine the number of observed events $n_{obs,i}$ for each bin in Eqs.~(\ref{eq:Li1},\ref{eq:Li},\ref{eq:likelihood}).  In Eqs.~(\ref{eq:Li1},\ref{eq:Li},\ref{eq:likelihood}), $S_i$ and $B_i$ are the same as used to set the mean and standard deviation for $n_{obs,i}$ sampling.  For each pseudoexperiment, we maximize the likelihood function Eq.~(\ref{eq:likelihood}) to find the best fit value for the signal strength and then determine the 68\% CL on $\mu_{Vh}$.  For each category, we average the best fit values and error bars over all pseudoexperiments.

In Fig.~\ref{fig:mu_cat} we show the signal strength mean value and 68\% CL for each of the 6 categories at 14 and 27 TeV with 3 ab$^{-1}$ and 15 ab$^{-1}$ of data, respectively.  As expected, all categories at a given lab frame energy $\sqrt{S}$ have the same central values for $\mu_{Vh}$.  Also, the two jet categories have much smaller uncertainties than the three jets, indicating little information is gained from the three jet categories.

\begin{figure}[tb]
\subfigure[]{\includegraphics[width=0.45\textwidth,clip]{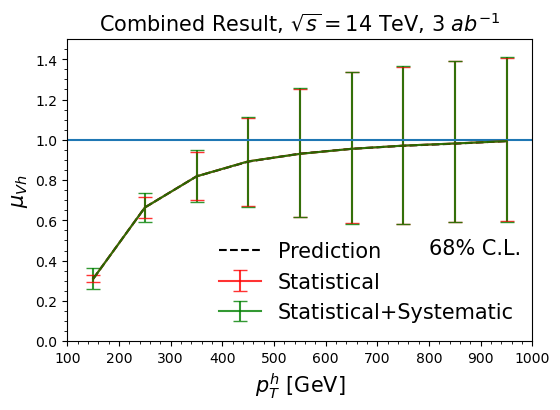}}
\subfigure[]{\includegraphics[width=0.45\textwidth,clip]{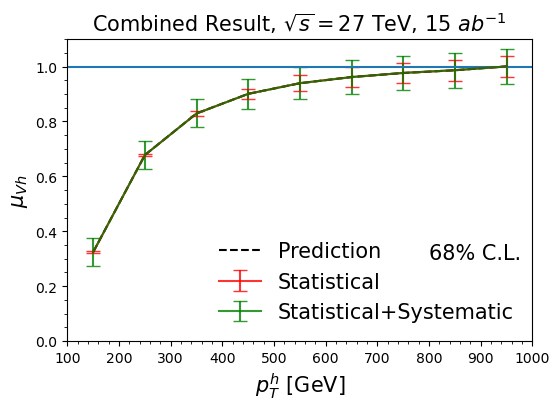}}
\caption{\label{fig:mu} Combined central values across all categories and 68\% CL for extracted signal strengths $\mu_{Vh}$ at (a) 14 TeV with 3 ab$^{-1}$ and (b) 27 TeV with 15 ab$^{-1}$. Black dashed lines are the partonic level prediction, the red bars are statistical uncertainty, the green bars are statistical and a 5\% systematic uncertainty added in quadrature.}
\end{figure}

Now that we have the individual signal strengths, we can combine them.  To do this, we create a ``global'' likelihood that is the product of the likelihoods for the six signal categories.  Then we perform the same procedure above with pseudoexperiments for each category to find the central value and 68\% CL for $\mu_{Vh}$.  These results are shown in Fig.~\ref{fig:mu} for both (a) the HL-LHC and (b) the HE-LHC, with the predicted partonic level signal strength overlaid.  As can be seen, the extracted central value is indistinguishable from the prediction.  In an optimistic scenario, the systematic uncertainty on $Vh$ production is expected to be 5\%~\cite{Cepeda:2019klc}.  The red uncertainty bands show the statistical uncertainty, and the green bands show statistical and a 5\% systematic uncertainty added in quadrature.

At low $p_T^h$, the signal strength is significantly far from one and then converges to one at higher energies, as expected.  Indeed, in the last overflow bin, we find the central value of the signal strength and 68\% CL to be:
\begin{eqnarray}
\mu_{Vh}=\begin{cases} 1\pm 0.4& \quad{\rm at~the~HL-LHC}\\ 1\pm 0.06& \quad{\rm at~the~HE-LHC}\end{cases}.
\end{eqnarray}
That is, the signal agrees with the EW restoration prediction at 40\% at the HL-LHC and 6\% at the HE-LHC. Hence, the $Vh$ rate converges to the expected rate with EW symmetry restored. This measurable convergence indicates empirically that the longitudinal modes can be replaced with the Goldstone bosons, and EW restoration can be observed at high energies.

\subsection{Statistical Test of EW Restoration}
To test how well EW restoration is being observed, one needs to measure how the convergence is improving by using higher and higher energy bins. At low $p_T^h$ bins, although the statistical error is small, the Goldstone and gauge boson distributions do not agree. As one moves toward higher $p_T^h$ bins, while the two distributions converge, the statistical errors also increase, as shown in Figs. \ref{fig:mu_cat} and \ref{fig:mu}.  In this section we explore statistical measures of the restoration and discuss their implications, taking into account both the theory convergence as well as the experimental uncertainties.  The goal is, assuming that the SM is a good description of the data, we want to test the agreement between the $q\bar{q}'\rightarrow Vh$ and $q\bar{q}'\rightarrow Gh$ ($\mu_{Vh}^j=1$) production as a function of $p_T^h$.

\begin{figure}[tb]
    \subfigure[]{\includegraphics[width=0.49\textwidth,clip]{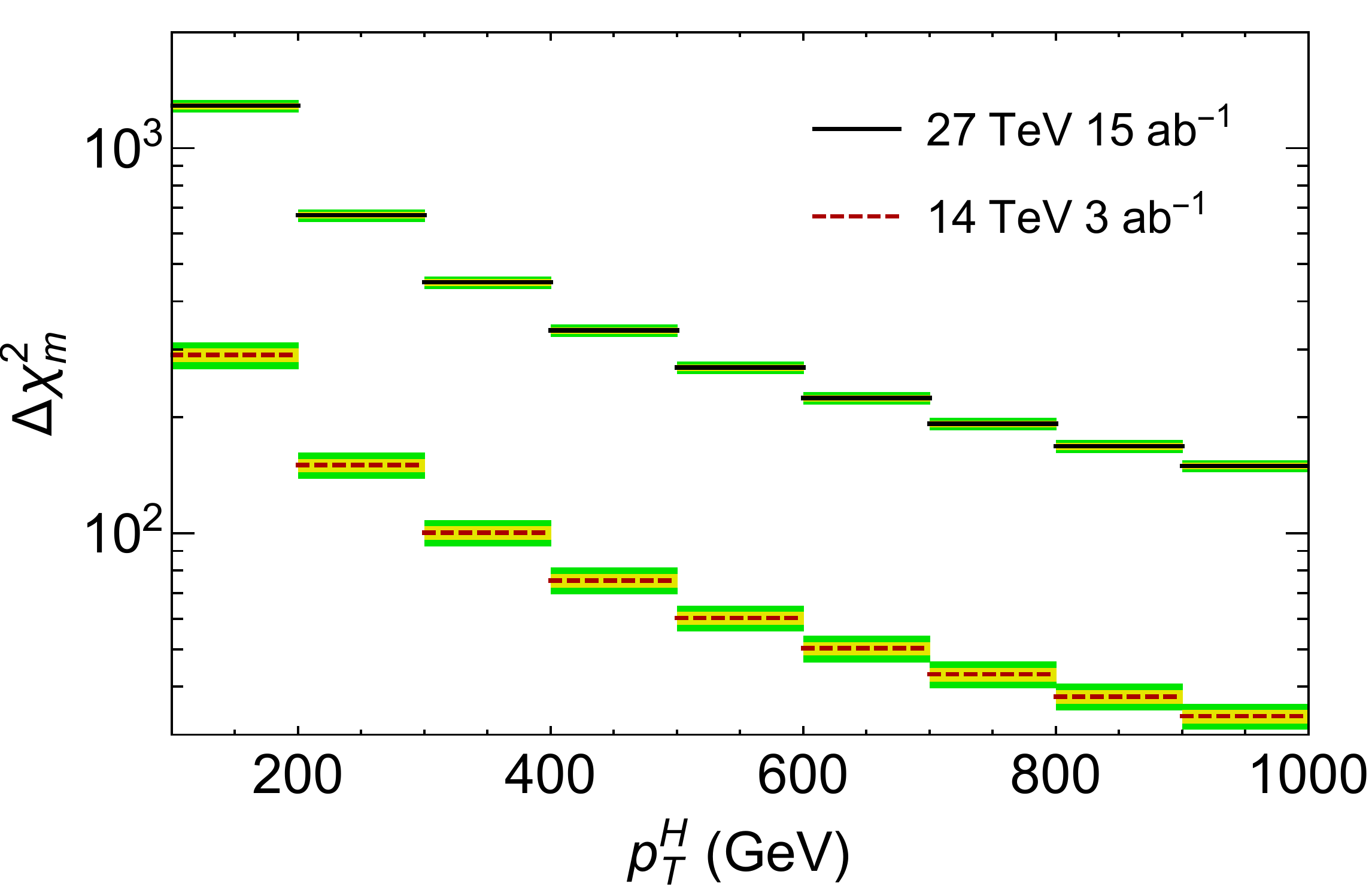}}
    \subfigure[]{\includegraphics[width=0.49\textwidth,clip]{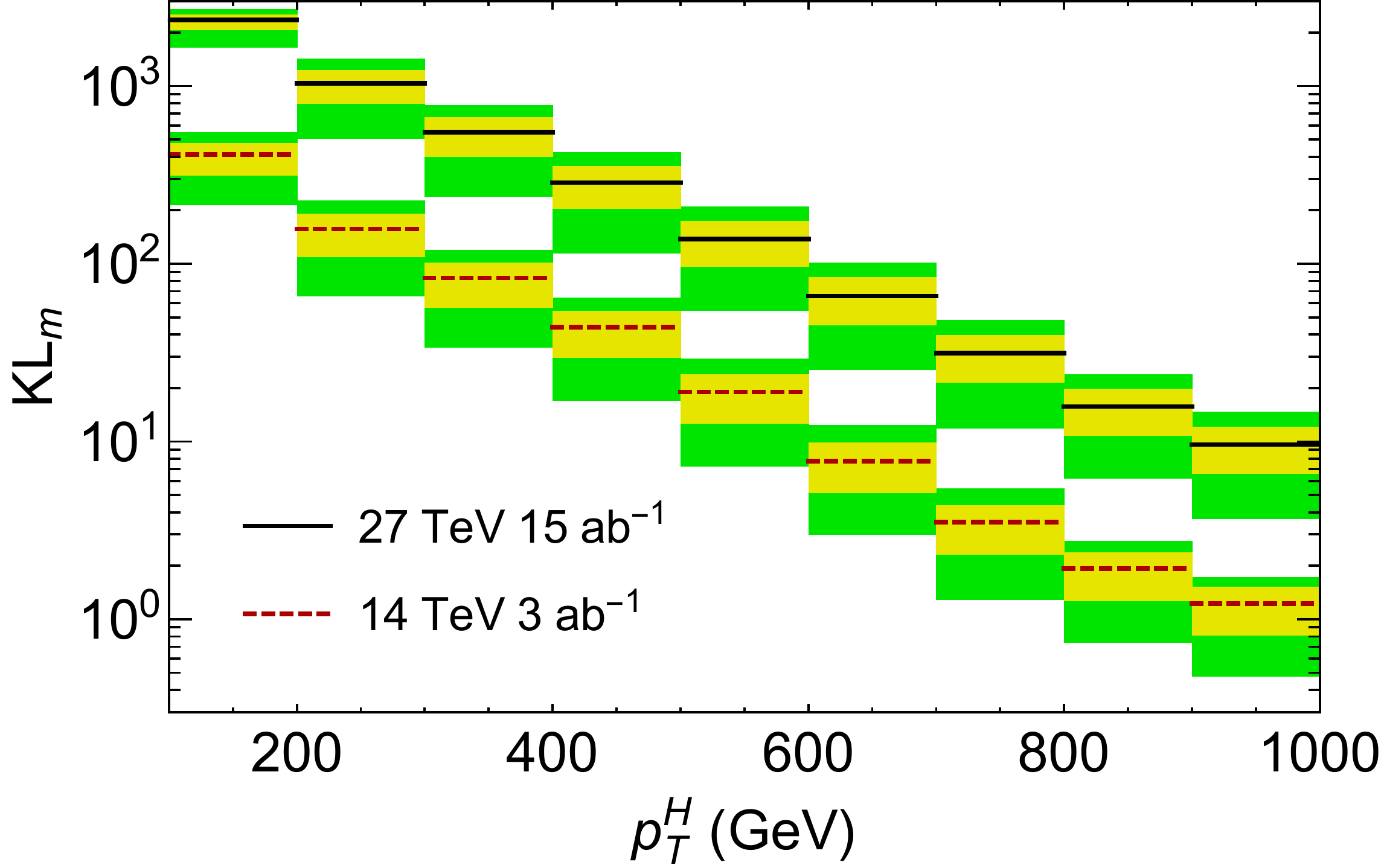}}
    \caption{\label{fig:KL} The $\chi^2$ per degree of freedom $\Delta \chi_m^2$ defined in Eq.~(\ref{eq:chi2dof}) (left panel) KL divergence defined in Eq.~\ref{eq:KL} (right panel) for 14 TeV with 3 ab$^{-1}$ and 27 TeV with 15 ab$^{-1}$. The back (red dashed) line represents the median values of the 27 TeV (14 TeV) results, and the yellow and green bands represent the values where 68\% and 95\% of pseudoexperiments lie, respectively.
    }
\end{figure}

As a first choice, using the language that the high energy physics community is more familiar with, we consider using ``$\chi^2$ per degree of freedom'' as a function of $p_T$ bins. One generically anticipates this quantity to decrease as an indicator of better convergence. 
After using the method in the previous sections in separating signal and background, we now have six-category samples, post-selection cuts, that have the 
significance of our analysis as a function of $p_T^h$. One can define ``$\chi^2$ per degree of freedom''\footnote{Here we use the log-likelihood ratio as delta chi-square for each bin.}:
\beq
\Delta\chi_{m}^2=\frac 1 m \sum_{l=1}^m \log\left(\frac {{\rm Pois}(n_{obs,l}|\sum_j \Delta \sigma_j^{Gh}\epsilon_{lj}L+B_l)} {{\rm Pois}(n_{obs,l}|S_l+B_l)} \right),
\label{eq:chi2dof}
\eeq
where we sum over the $m$ ranked $p_T^h$ bins (from low to high).  Using the methods of the previous section, we perform 10,000 pseudoexperiments. 
 The results are shown in the left panel of Fig.~\ref{fig:KL}.  We show the median over all pseudoexperiments as well as the band where 68\% and 95\% of the pseudoexperiments lie. From the figure we can see, as anticipated, the $\Delta \chi^2_{m}$ decreases as one includes more high $p_T^h$ bins.   

However, we note here that $\Delta\chi_{m}^2$ has some disadvantages in measuring restoration. First, for the low $p_T^h$ bins, each bin contributes to a sizable $\Delta\chi^2$ since the $Gh$ and $Vh$ hypothesis are in poor agreement and statistical uncertainty is small. At high $p_T^h$, the statistical uncertainties increase.  Hence, even if the $Gh$ and $Vh$ distributions do not converge, as more bins are averaged over $\Delta \chi^2_m$ will decrease.  
In other words, even if the higher bins contain no separation power, e.g. the background uncertainty being infinitely larger than the signal strength, the $\Delta\chi_{m}^2$ decreases.  This reflects that $\Delta\chi_m^2$ measures the agreement between two hypotheses: as the uncertainties increase, the error bars overlap, and the hypotheses are in ``good agreement.''  However, to measure EW restoration, the convergence of $Vh$ and $Gh$ must be measured and $\Delta\chi_m^2$ is not a good measure of convergence. 

As can be seen, the measurement of the EW restoration is not a typical particle physics test.  The issue is that we want to measure the convergence of two hypotheses with energy, not just determine how well they agree globally.  
Ideally, the measure should contrast different hypotheses for a given experimental data set with proper weight for each bin according to the ``information'' contained there. We turn to Shannon's information theory and find that generically $-p\log p$ measures the information of a distribution $p$. 
While there might be an equivalent or better definition outside of our scope, we use a modified Kullback-Leibler (KL) divergence.  
KL divergence is a commonly used quantity contrasting the information between two different distributions, and often plays the role of loss function for machine learning. 
The KL divergence tests the information difference between two hypotheses.  To do this, for each pseudoexperiment we first define properly normalized probability for each bin for the $Vh$ hypothesis 
\begin{eqnarray}
p_i^{\leq m}=\prod_{\substack{{\rm 6\, signal}\\{\rm categories}}}\frac{{\rm Pois}(n_{obs,i}|S_i+B_i)}{\sum_{l=1}^m {\rm Pois}(n_{obs,l}|S_l+B_l)},
\end{eqnarray}
where the $i\leq m$ are bin numbers with increasing $p_T^h$.  We have assumed independent event samples, and so have taken a product of probabilities across all signal categories. Restricting ourselves to $p_T^h\leq p_{T,m}^h$ with $p_{T,m}^h$ being the central value of $p_T^h$ in the $m$th bin, $p_i^{\leq m}$ is the probability of observing $n_{obs,i}$ events in bin $i$ given a SM hypothesis of $S_i+B_i$ bins.  The $Gh$ hypothesis is equivalent to signal strengths of one: $\mu_{Vh}=1$.  Using the efficiency matrices $\epsilon_{ij}$ we can define an analogue conditional probability for the $Gh$ hypothesis:
\begin{eqnarray}
q_i^{\leq m}=\prod_{\substack{{\rm 6\, signal}\\{\rm categories}}}\frac{{\rm Pois}(n_{obs,i}|\sum_j \Delta \sigma_j^{Gh}\epsilon_{ij}L+B_i)}{\sum_{l=1}^m {\rm Pois}(n_{obs,l}|\sum_j \Delta \sigma_j^{Gh}\epsilon_{lj}L+B_l)},
\end{eqnarray}
where the sum over $j$ is over all bins and {\it not} restricted to bins less than $p_{T,m}^h$.  The KL divergence for the first $m$ bins is then:
\begin{eqnarray}
KL_m= \sum_{i=1}^m p_i^{\leq m}\,\log\left(\frac{p_i^{\leq m}}{q_i^{\leq m}}\right).\label{eq:KL}
\end{eqnarray}
Now the interpretation of the KL-divergence is clear.  If the two hypotheses describe the data equally well, the $\log$ goes to zero and the KL divergence is zero. The KL-divergence has a similar property as the Gibbs free energy, being positive definite.  Hence, when the agreement of the two hypotheses is worse, $KL_m$ is larger.  As more bins are included, we expect the EW restoration to describe data better and the KL divergence should approach zero.

When the two hypotheses do not agree, the weighted sum in Eq.~(\ref{eq:KL}) guarantees that the largest contributions come from bins for the conditional probabilities $p_i^{\leq m}$ are largest.  Hence, the KL divergence contains more information than $\Delta\chi_m^2$ and is expected to be a better measure of convergence.  
In Fig.~\ref{fig:KL} we show the differential KL divergence, $KL_m$.  We show the median over all pseudoexperiments as well as the band where 68\% and 95\% of the pseudoexperiments lie.  As can be clearly seen, whereas the $\chi^2$ per degree of freedom test began to plateau at high energies, the KL-divergence decrease more steadily.  This more readily shows that the agreement of the $Vh$ and $Gh$ hypotheses continues to get better at high $p_T^h$ and we observe EW restoration.

We want to emphasize here that the convergence between $Vh$ and $Gh$ distributions is directly represented by the fact that $\Delta\chi_m^2$ and $KL_m$ decrease as higher and higher $p_T^h$ bins are included.  
We would like to note that somewhat counterintuitively the 14 TeV statistical tests seem to be ``better'' than the 27 TeV results.  That is, the 14 TeV values are lower.  Even if we had $p_T$ bins larger than 1 TeV, the 27 TeV results will not ``beat'' the 14 TeV results.  This is because the smaller uncertainties at 27 TeV cause the first bin, where agreement is poor, to be considerably greater than the 14 TeV results.  This results in the entire 27 TeV $\Delta\chi_m^2$ and $KL_m$ distributions being greater than at 14 TeV.  That is, although the uncertainties of the differential cross sections between different machines can be compared to determine which machine is more sensitive, the nonconventional tests of the convergence of the signal strengths, i.e. $\Delta\chi^2_m$ and $KL_m$, should be considered on a machine-by-machine basis.

\section{Conclusions}
\label{sec:conc}
In this paper, we have studied the potential of the HL-LHC and HE-LHC to observe EW restoration in $pp\rightarrow Vh$ production. Discussions of EW symmetry restoration have traditionally been limited to the longitudinal vector boson scattering. Using the Goldstone boson equivalence theorem, it can be seen that this scattering occurs via the quartic term in the Higgs potential. However, Goldstone bosons also have interactions via the Higgs kinetic terms. These terms contribute to the production of longitudinal gauge bosons in $q\bar{q}'\rightarrow VV'$ and $q\bar{q}'\rightarrow Vh$ channels.

As we showed, the $q\bar{q}'\rightarrow VV'$ production is dominated by transverse polarizations to very high energies. Hence, it is difficult to observe Goldstone boson production in this channel. Since $q\bar{q}'\rightarrow Vh$ is a purely $s$-channel process with a component of the Higgs doublet, it is longitudinally dominated starting at relatively low energies. In Sec.~\ref{sec:Theory}, we defined EW symmetry restoration by taking the limit of the Higgs vev going to zero and enforcing the SM tree-level relations for the Higgs potential parameters. This results in a massless Higgs doublet in the EW restored theory. From this, we defined a differential signal strength $\mu_{Vh}$ as the ratio of the $p_T^h$ distributions of $Vh$ and $Gh$ production. As shown, this signal strength is the same for $Zh$ and $W^\pm h$, allowing for easy extraction of global signal strength in all $Vh$ channels. This convergence can be seen in Fig.~\ref{fig:sigstrength}.

As EW symmetry is restored, the longitudinal gauge bosons are replaced with their Goldstone boson counterparts. Hence, the signal strength $\mu_{Vh}$ is expected to converge to the Goldstone calculation at high energies. Using a sophisticated collider analysis, we showed that by performing a fit to these signal strengths, it can be observed that the $Vh$ channel converges to $Gh$ at high energies. Indeed, for $p_T^h\sim 400$~GeV, the $Vh$ and $Gh$ distributions agree at around 80\%. Additionally, as can be seen in Figs.~\ref{fig:mu_cat} and~\ref{fig:mu}, the extracted signal strength in all $Vh$ signal categories agrees with the partonic level prediction. Finally, to quantify the agreement between the $Vh$ and $Gh$ hypothesis, we defined a differential Kullback-Leibler divergence. If two hypotheses agree, the KL divergence is small. As shown in Fig.~\ref{fig:KL}, the $Vh$ and $Gh$ hypothesis agree well at high energy, and EW restoration can be well observed.

To summarize, we demonstrated the EW restoration could be observed in the $Vh$ channel. Indeed, EW restoration can be confirmed to 40\% precision at the HL-LHC and 6\% precision at the HE-LHC. Our study can be further extended to other future colliders, as well as the other di-boson production modes highlighted in the theory discussion. 
Our study clearly highlights the possibility of studying the physics phenomena of electroweak restoration at high energy colliders as well as electroweak breaking.


\section*{Acknowledgments}
I.M.L. would like to thank the Institute for Theoretical Physics at Universit\"{a}t Heidelberg for their hospitality during the beginning of this work and Tilman Plehn for insightful discussions.  I.M.L. would also like to thank Sally Dawson for useful comments.  This work was performed in part at the Aspen Center for Physics, which is supported by National Science Foundation grant PHY1607611.  L.H. and I.M.L. were supported in part in part by United States Department of Energy grant number de-sc0017988. SL is supported by the State of Kansas EPSCoR grant program and the U.S. Department of Energy, Office of Science, Office of Workforce Development for Teachers and Scientists, Office of Science Graduate Student Research (SCGSR) program. The SCGSR program is administered by the Oak Ridge Institute for Science and Education (ORISE) for the DOE. ORISE is managed by ORAU under contract number DE-SC0014664. ZL was supported in part by the NSF grants PHY-1620074, PHY-1914480 and PHY-1914731, and by the Maryland Center for Fundamental Physics (MCFP).  The data to reproduce the plots has been uploaded with the arXiv submission or is available upon request.

\appendix
\section{DNN Inputs}
\label{app:DNN}
First, note the Higgs is always reconstructed from the two leading bottom tagged jets. In the 2-lepton categories the $Z$ is reconstructed from the two leptons, and in the zero lepton category the $Z$ is reconstructed from the missing transverse energy.  Note for the zero lepton category, we only reconstruct the $p_T$ of the $Z$.  In the 1-lepton case, there is a missing neutrino.  Its transverse momentum is assumed to be the missing transverse energy of the event.  The neutrino's longitudinal momentum is found by requiring that the neutrino plus lepton system reconstruct the $W$-mass.  This leads to a two-fold ambiguity, and we choose the neutrino momentum that is closer to the lepton.  Hence, in the 1-lepton case the $W$ is reconstructed.  Additionally, we label the leading bottom jet at $b_0$, the next to leading bottom jet as $b_1$, and the leptons as $\ell_0,\ell_1$ similarly.  For 3-jet categories, we label the leading non-$b$ jet as $j$.  

The following definitions are used:
\begin{itemize}
\item  The invariant mass of two objects $i,j$ is $M_{ij}$.
\item The reconstructed mass of an object $i$ is $M^{recon}_i$.
\item For a final state system $i$, the transverse mass is defined at $M_{T,i}=\sqrt{E_i^2-p_{Z,i}^2}$, where $E_i$ is the total energy of the objects $i$ and $p_{Z,i}$ is the $z$-component of their momentum.
\item The transverse momentum of an object $i$ is $p_T^i$.
\item The azimuthal difference between two objects $i,j$ is $\Delta \phi_{ij}=|\phi_i-\phi_j|$.
\item The difference between the rapidities of two objects $i,j$ is $\Delta \eta_{ij}=|\eta_i-\eta_j|$.
\item The opening angle between two objects $i,j$ is $(\Delta R_{ij})^2=(\Delta\phi_{ij})^2+(\Delta \eta_{ij})^2$.
\item The scalar sum of all transverse momentum is $H_T$.
\item The number of non-$b$ jets is $n_j$.
\end{itemize}

 The input variables for the DNN for each event category are:
\begin{itemize}
\item 2-lepton, 2-jet:
\begin{itemize}
\item $M_{Zh}$, $M^{recon}_h$, $M^{recon}_Z$.
\item MET, $p_T^h$, $p_T^Z$,  $p_T^{b_0}$, $p_T^{b_1}$, $p_T^{\ell_0}$, $p_T^{\ell_1}$, $H_T$.
\item $\Delta \phi_{Zh}$, $\Delta \eta_{Zh}$, $\Delta R_{Zh}$.
\item $\Delta R_{b_0 b_1}$, $\Delta R_{\ell_0 \ell_1}$, $\Delta R_{b_0 \ell_0}$, $\Delta R_{b_1 \ell_1}$.
\item Transverse mass of reconstructed Higgs and Z, $M_{T,Zh}$.
\end{itemize}
\item 2-lepton, 3-jet: Same as 2-lep+2-jet with:
\begin{itemize}
\item $p_T^j$, $\Delta R_{hj}$, $\Delta R_{Zj}$.
\item $n_j$.
\end{itemize}
\item 1-lepton, 2-jet: 
\begin{itemize}
\item $M_{WH}$, $M_h^{recon}$, $M_W^{recon}$.
\item MET, $p_T^h$, $p_T^W$, $p_T^{b_0}$, $p_T^{b_1}$, $H_T$. 
\item $\Delta \phi_{Wh}$, $\Delta R_{b0 b1}$, min\{$\Delta \phi_{\ell 0 b 0}$, $\Delta \phi_{\ell 0 b 1}$\}.
\item Transverse mass of reconstructed Higgs and W, $M_{T,Wh}$.
\item Transverse mass of $W$: $M_{T,W}$. 
\item Transverse mass of $W+b_0$: $M_{T,W b_0}$. 
\item Transverse mass of $W+b_1$: $M_{T,W b_1}$. 
\item $|\Delta Y_{hW}|=|\eta_{h}-\eta_{W}|$, where $\eta_{h,W}$ are the Higgs and $W$ rapidities.
\end{itemize}
\item 1-lepton, 3-jet: Same as 1-lepton, 2-jet with:
\begin{itemize}
\item $p_T^j$, $M_{hj}$
\end{itemize}
\item 0-lepton, 2-jet: 
\begin{itemize}
\item $M_h^{recon}$.
\item MET, $p_T^h$, $p_T^{b_0}$, $p_T^{b_1}$, $H_T$.
\item $\Delta \phi_{Zh}$, $\Delta\eta_{b_0 b_1}$, $\Delta R_{b_0 b_1}$.
\item Transverse momentum of reconstructed Higgs and Z, $M_{T,Zh}$.
\end{itemize}
\item 0-lepton, 3-jet: Same as 0-lepton, 2-jet with:
\begin{itemize}
\item $M_{hj}$, $p_T^j$.
\end{itemize}
\end{itemize}




\clearpage
\bibliographystyle{myutphys}
\bibliography{EWRestor}

\end{document}